# AN EVIDENCE FOR SOLAR ACTIVITY INFLUENCE ON THE METEOROLOGICAL PROCESSES IN THE SOUTH POLAR REGION OF MARS DURING THE GREAT OPPOSITION IN AD 1924

Boris Komitov

*Bulgarian Academy of Sciences – Institute of Astronomy*
*Bulgaria, 6003 Stara Zagora-3, POBox 39*
*b_komitov@sz.inetg.bg*

A time series of the Martian south ice polar cap mean diameter for the period July-December 1924 is investigated. The data are based on the high quality pictures , which are obtained by visual observations of 60-cm telescope in Hamburg Observatory during the great opposition of Mars in AD 1924. After removing of the seasonal trend (caused by the springtime regression of the cap) quasi 36 and 80-82 –days cycles in residuals has been obtained. The sunspot activity spectra for the corresponding period is almost the same one. The local maximums of polar cap area residuals has been occured ~ 10 days after the corresponding minimums of sunspot activity. The so obtained results are briefly discussed.

*1 . Introduction*

One of the most interesting and discutabile problems concerning the Earth's climate is about the role of solar activity for the changes of the last one. According the present point of view the climate of our planet depend by large number of factors, which influence is very dificult to separate each from other. By this fact , the study for climatic changes of other planets in Solar system , the searching for relationships with the solar activity may give an impotrant additional contribution to solving of the above mentioned problem.

For this aim Mars is the most suitable object. The planet is observed more than 300 years and since the 60's it is investigated by using of space probe based instruments. The processes of building and regression of polar caps , the dust storms , the generating and evolution of clouds and aerosol structures in atmosphere, the seasonal variations of color and albedo in separate regions of the surface are giiving an important information about the metheorology and climate of this planet.

Close to this moment the most certain indicator of Martian climatic changes are the polar caps dynamic data. They are observed from the 17$^{th}$ century. At the beginning of 20$^{th}$ century there are already many pictures on the base of visual observations, where has been provided by many researchers in differrent times. During the last ~100 years a large number high precission photographic and micrometric observations has been added. During the last two decades the "weight center" is already shifted to CCD-camera observations /ground based , from space probe boards as well as from the Hubble Space Telescope.

Since 1997 the the space probe orbiter modules observations are already practically continuous.

This is yet not enough for study of processes , where are related to the polar regions climate changes. Although the observations from Martian satelites shows clear evidences, that the Martian climate and especially in near polar regions is subjected of significant changes. For example the layer structure of the summer residual (water) north polar cap most probably is caused by different regression rates and, connsequently, by interleaving of warmer and coulder climatic epochs.

However the ground based telescopic observations in 20[th] century for climate called polar caps regression rate changes remain the main important information source on this stage.

Antoniadi [1] is the first researcher, who conclude that the polar caps regression rates are depended by solar activity level. On the base of visual drawings analysis for the period 1852-1912 he established that the polar caps regression rates are smaller during the years of solar sunspot activity minimums as in the years of maximums. Bassu make similar conclusions [2]. He analysed the observation data /mainly photographic inmages/ for the period of 1905-1988. However the observational material which has been used in the both abovesaided studies is very unhomogenious. The differrences are caused mainly by the different methods, telsescopes, individual features and experience of observers as well as by the differrent conditios of Mars visibility. Some phenomenas such as dust storms, aerosols or clouds in polar regions can be load to significant errors. *Although the conclusion about the revert dependence between the solar activity level and polar caps regression rates seems to be logical, because the coulder climate /slow regression/ must corresponding to lower solar radiation levels.

Obviously the solution of this problem is on its initial stage. The using of larger and larger number of observations from orbiting arround Mars space probes will play the most important role in the near future.

## 2. The aim of present study. Data and initial proceedings

A typical feature of ground based observations of Mars during the 19[th] and 20[th] centuries is in the fact, that they are very intensive when the so called "great oppositions"occurs. With most higher probability during these periods is possible to obtain series of drawings and photographies from one and the same observers, by using of one and the same telescopes, observational technique and image proceeding trough short time intervals between separate observations and during of few months periods.

It could be used such image series for metheorological conditions tracking over chosen regions of the planet during the period of observations. If (eventually) short-time quasi-cyclic variations will be obtained, a comparison with the short – time variations of solar activity will be reasonable. An other aim in this course can be determination of the "resident time" of the solar activity level changes influence over processes in near-surface atmosphere.

Such homogeneous series, containing 38 qualitative drawing images of Mars has been obtained by Kazimir Graff in 1924 by using of 60-centimeter refractor in Hamburg observatory[3] (fig.1). The observations has been provided during the period July-December of the same year. The south polar cap is clear shown. The period of observation contain the end of winter and the beginning of spring in the south Martian hemisphere. The last one give for Graff the possibility for tracking of the mean angular diameter $d$ of south polar cap during the period of its melting.

On the base of smoothing data curve the values of $D$ has been determined by step of 5 days. The last one correspond approximately to the mean time interval between two adjacent observations. The so obtained time series is an object of our next analysis.

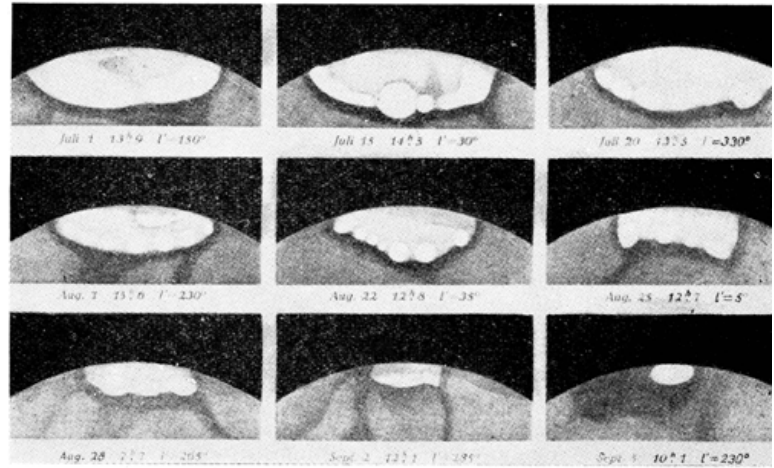

*Fig1. Drawings of the South polar cap melting in 1924 by K.Graff (published in [3])*

The daily smoothed sunspot activity index *Ri* /the International Wolf's number/ for the period 1818-2004 could be downloaded by Internet at address: *ftp://ftp.ngdc.noaa.gov/STP/SOLAR_DATA/SUNSPOT_NUMBERS*. The part of values, which correspond to the Graff's period of observations has been averaging for 5 days. Thus the so obtained time series of sunspot activity correspond to the mean south polar cap angle diameter series.

The both derived final modified time series of *Ri* and *D* has been analysed for existing of trends and cycles.

*3. The T-R periodogramm analysis*

In the present work a numerical procedure, called "T-R periodogramm analysis" (TRPA) for detection of cycles in time series has been used. Comparing with the "classical" Fourier-analysis and the large part from other usualy used methods, it is essentialy more comfortable and sensitive for detection of cycles when the ratio *N/T* is non-integer value in the common case. (*N* is the time series length and T- the period, which corresponded to the length of cycle). For first time TRPA is described in [ 4]. A more detailed and developed version has given in [5]. It contain the follwing steps:

1. An approximation of the investigated time series by using of the least square procedure over a series of simple periodic functions :

$$f(t) = Ao + A \cos(2\pi t/T) + B \sin(2\pi t/T) \quad (1)$$

where *Ao* is the mean value of all terms in time series and the coefficients *A* and *B* can be determinate by solution of the system:

$$\sum_{i=1}^{N} (y_i - Ao)\cos\frac{2\pi(i-1)}{T} = A\sum_{i=1}^{N}\cos^2\frac{2\pi(i-1)}{T} + B\sum_{i=1}^{N}\sin\frac{2\pi(i-1)}{T}\cos\frac{2\pi(i-1)}{T}$$

$$\sum_{i=1}^{N} (y_i - Ao)\sin\frac{2\pi(i-1)}{T} = A\sum_{i=1}^{N}\cos\frac{2\pi(i-1)}{T}\sin\frac{2\pi(i-1)}{T} + B\sum_{i=1}^{N}\sin^2\frac{2\pi(i-1)}{T}$$

(2)

where $y_i = F(t)$ are the terms of the time series and the corresponding moments are $t = 0, 1, 2, \ldots N - 1$. The time interval of "1" between the adjacent terms is the "time series unit", i.e. the step of the time series. It real value may be equal of year, day, minute e.t.c. /In our case this step is equal of 5 days./

The period T is varied by equidistant step $\Delta T$ from choosed value $T_0$ to some maximal one $T_{max}$. The lower possible limit for $T_0$ is 2.

$$SR = \frac{1 - R^2}{\sqrt{N}}$$

2. For every one of the so obtained simple periodic minimized functions the corresponding coefficient of correllation $R$ with the time series, as well as its error are calculated. The obtained series of $R$ values (T-R –corellogram) have local maximums near this values of $T$, where corresponded of potentially existing cycles in the time series. The amplitude (power) of the cycle may calculated by formula:

$$a(T) = \sqrt{A^2(T) + B^2(T)}$$

(3)

3. Tests for statistical validity of the cycles. A two criteria are used there. According the first one it need $R/SR > 1.96$. However very often in pseudo-random number series are generated cycles, where the abovesaid test is covered. Because of this fact in [5] a second, more hard test is oferred. It has been established on the base of analysis of more than 5000 pseudo-random number series. According the last one it needed $R/SR > 4.54/N^2 + 3.5$ for "non-random" generating of cycle. It is clear that for enough long series ($N \to \infty$) the "critical" value of $R$ tend to 3.5. In the cases when the $R$ value is between the both "critical" limits the problem about the cycle validity is solved on the base of other expert estimations.

4. Quasi 36-37 and 80-days oscilations of mean temperatures in south polar region of Mars during the second half of 1924[th]

As it shown on fig.2. the time sereis of D /circles/ can be separate on three parts:
1. The most left presented the dynamic of south polar cap dimension in July 1924. It characterized by slight decreasing of D, i.e. monotonic melting of the polar cap.

2. A phase of fast regression /melting/ which started aproximately at the beginning of August . This process is not light and two "waves" of temporal increasing of $D$, i.e. recovering of the ice cap are shown.

3. Near to end of November the regression has been practically stoped and the mean polar cap diameter $D$ remain near to 10 °.

By using of a regressional procedure tests for choosing of the best fit function for decribing of the polar cap diameter dynamic as non-linear trend has been made. It turned out, that the best expression of the trend is a full polynom of fourth degree (the flat line on fig 2). It is clear shown that the trend line practicaly coincides with the time series data in regions "1" and "3". The quasi-cyclic deviations from the trend are very clear expressed in phase "2". Phase "3" corresponds of non-volatile residual /dust mantle/ formation. This stop the further sublimation of "dry" ice ($CO_2$) as well as the destroying of polar cap too.

This behaviour of $D$ is in good agreement with hypotesis , that in the reached Martian surface solar radiation exist a component , which by one hand is strong absorbed at large solar zenith angles and by other it caused strong quasi-cyclic variations with near-month duration. The solar UV-radiation at $\lambda \leq 180$ nm is satisfied to the both features. However its absorbtion in the low Martian atmosphere is very strong and it almost not reached to the surface. Consequently ,it remain only the possibility that the abovesaid quasi-cyclic solar component is the high energetic corpuscilar radiation. As is well known the basical near-monthly cycle of the observed solar activity is by duration of 26-28 days during the decreasing phases of 11-year solar cycles. It is almost one and the same by observations both from Earth and Mars . However, during the increasing phases very often the quasi 28-28 days cycle is weaker relatively to oscilations with duration larger than mont

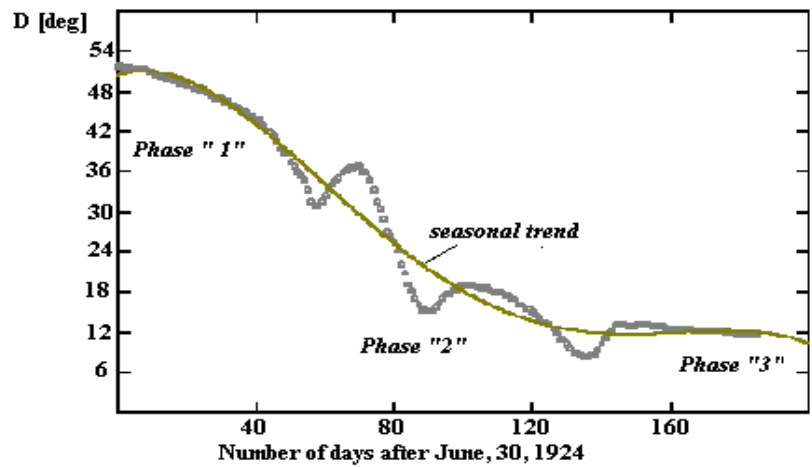

*Fig.2 The mean angular diameter D of Martian south polar cap (by dark circles) (Jul-Dec, 1924) and the seasonal trend (the flat line) , which is approxymated by a full polynom of fourth degree.*

That is why to be verify the hypothesis of the solar determining of the mean diameter variations for the south polar cap in the phase "2" - an analysis has been made on the base of the following steps :

1. Removing of the non – linear trend from the time series of the values concerning $D$.

2. Removing from the new established series the residual variations of the values, refer to the phases "1" and "3".

3. The reduction time series of the residual variations $d$, related to the phase "2" and containing totally 21 values / for the period of time 110 days / has been investigated for the presence of statistical reliable cycles by means of T-R periodogramm analysis [4].

4. The so obtained T-R corellogram for the series of the residual variations of the polar cap mean diameter has been comparing with the corresponding one, but derived for the time series of the avarage for five days values of the sunspot index $Ri$ for the same time.

The results are shown on Figs. 3 and 4. It is evidently that at the first of them in the variations of the mean diameter of the south polar cap, during the period August – November 1924, presents a good showed 36 –day cycle as well as and a second one ,weaker – but reliable quasi 80 - days cycle .The last one is resonancely multiple of the "classical" 27 – solar cycle.

It is clear shown from fig.4 that the main near-montlhly oscilation of solar activity level is by duration of 36-37 days. It is almost equal to the basic cycle of the polar cap variations. The traces of 27-days cycle are weak, but in addition there is very clear visible and statistical valid quasi –80-days oscilation in sunspot activity. Consequently, the both correllograms in fig.3 and 4 are very similar. This indicate that a strong relationship /may be anti-corellation/ between $d$ and $Ri$ may to expect.

The searching for direct relationship between these both parameters point out, that the maximal by module correllation coefficient $r = -0.73$ occur when the phase shifting is 10 days . $Ri$ overtackes $d$, i.e. the maximal polar cap rebuilding occur about 10 days after the local minimum of the sunspots. The statistical validity of this relationship exceed 99.9%.

The physical explanation of this phenomena could be search in two courses:

1. The penetrating to the surface solar variable component /protons by energy >1 MeV/ reach local maximums near to the maximal levels of sunspot activity. The last one lied to corresponding changes of near surface atmosphere increasing the aerosol production rate and consequently- to temperature and polar cap regression rate variations. The "resident time" of this process is about 10 days. The observed cases of particular polar cap rebuilding are corresponding to maximal cooling of atmosphere. Then the $CO_2$ "dry ice" sublimation rate is going down and the condensation is more active.

2. During the near sunspot minimum conditions as in AD 1924 an other mechanism could be much more probable. It is related to the increasing of the falling in Martian atmosphere galactic cosmic rays (GCR) flux. The higher values of the last ones corresponds to low level of the solar wind flux parameters (the Forbush -effect). Thus if any solar active center is going out of view from the Mars possition it should be follow an decreasing of solar wind and increasing of GCR flux near to the planet 3-4 days later. As in the case "1" the GCR flux increasing should to intensifed the aerosol production rates . This mechanism has been firstly described by Svensmark and Friis- Cristiensen [6]. It should also taken into account also a possible "resident time" which need to generate a meteorological effect of cooling in the atmosphere.

It should be note , that there is also an uncertainity, which come from the different observational conditions of the Sun from the Earth and Mars. While the published *Ri* data are Earth- related , i.e the observational conditions from the Mars position in one and the same moments should be different. It concern the solar wind and GCR –flux conditions in the Martian near space too. However the possible differnce by the above mentioned geometrical causes should be not very large, because of the fact of the Mars opossition in the second half of 1924[th]. On other hand it is absolutely possible that a significant part of the 10 days shifting in the relationship between *Ri* and *d* has been caused namely by this geometrical factor. Thus by the observation geometry as well as by the physical causes which are described in "1" and "2" it is very difficult to discussed of the relationship shifting in some certain physical aspect. It is only an observational phenomena in this study.

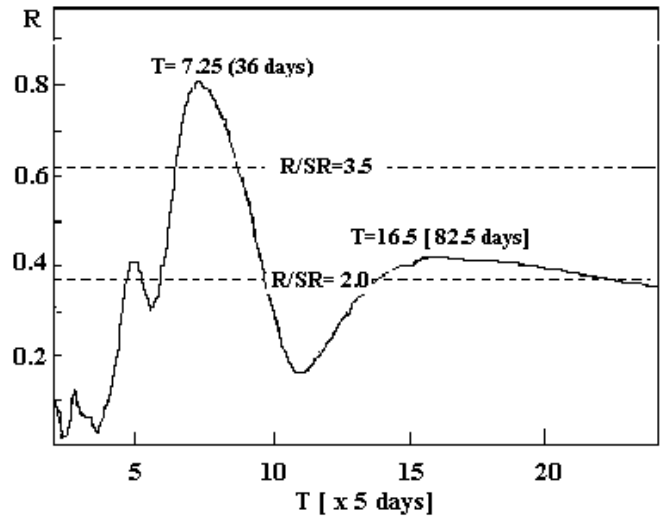

*Fig 3. The T-R corellogram of the d –residuals (August 1 –November 20, 1924)*

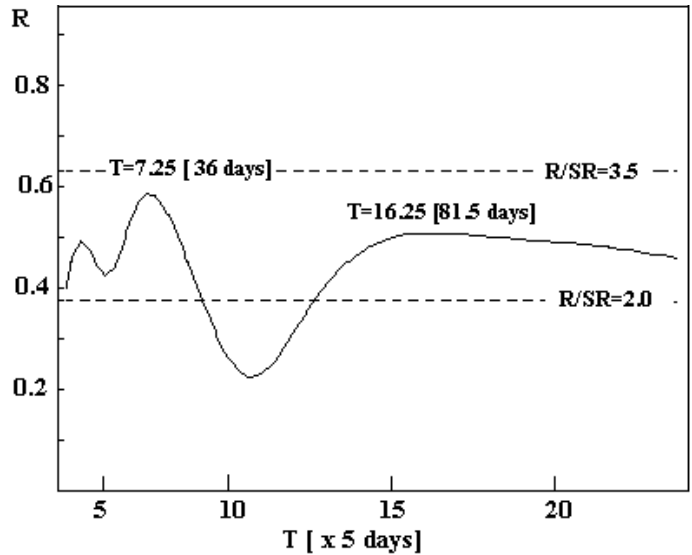

*Fig 4. The T-R corellogram of the dayly Inyernational Sunspot Number (Ri) series (August 1 –November 20, 1924)*

**5. Conclusion**

The obtained results shown that homogeneous photographic images series, obtained during the "great oppositions" in 20[th] century in relatively short time intervals / few days/ can be used for study of short periodic metheorological parameters oscilations in choosen regions of Mars, such as the polar caps. It is evidently of the all things that, the role of acive processes on the Sun over the meteorology and climate of this planet is significant.

It can be continued the investigation by using of additional photographic matherial from ground –based observations, but first of all- on the base of space probes information.